\documentclass[a4paper,12pt,english,german]{article}

\begin{document}

\centerline{\bf Asymptotic dynamics of the alternate degrees of
freedom}

\centerline{\bf for a   two-mode system: an analytically solvable
model\footnote{The work on this paper is financially supported by
Ministry of Science Serbia under grant no 171028.}}

\bigskip

M. Arsenijevi\' c$^{\ast}$\footnote{Email: fajnman@gmail.com}, J.
Jekni\' c-Dugi\' c$^{\dag}$, M. Dugi\' c$^{\ast}$

\bigskip

$^{\ast}$Department of Physics, Faculty of Science, Kragujevac,
Serbia

$^{\dag}$Department of Physics, Faculty of Science and
Mathematics, Ni\v s, Serbia

\bigskip

The composite systems can be non-uniquely decomposed into parts
(subsystems). Not all decompositions (structures) of a composite
system are equally physically relevant. In this paper we answer on
theoretical ground why it may be so. We consider a pair of
mutually un-coupled modes in the phase space representation that
are subjected to the independent quantum amplitude damping
channels. By investigating  asymptotic dynamics of the degrees of
freedom, we find that the environment is responsible for the
structures non-equivalence. Only one structure is distinguished by
both locality of the environmental influence on its subsystems and
a classical-like description.

\bigskip



\bigskip

{\bf 1. Introduction}

\bigskip

Realistic physical systems are composite--i.e. decomposable into
smaller "parts" (subsystems). The set of the "subsystems" (i.e. of
the degrees of freedom) of a composite system is {\it not} unique.

In classical physics, only one such set of subsystems (e.g. of the
constituent particles) is usually considered physically relevant.
The alternate decompositions (structures) of the composite system
are typically considered non-realistic, a mathematical artifact.
However, in the quantum mechanical context, the things may look
different.

There is ongoing progress in distinguishing physical relevance of
the alternate structures of a composite quantum system both on the
foundational as well as on the level of application, cf. e.g.
Refs. [1-12]. Regarding the {\it foundational} issues, the
following question is of interest: which degrees of freedom of a
composite system provide the above-mentioned classical description
[2, 3, 8, 9, 11, 12]? A closely related {\it interpretational}
question reads: is there a physically fundamental set of the
degrees of freedom of a composite system [2, 3, 10]? In the
context of physical {\it application}, one can differently
manipulate the different structures of a composite system, e.g.,
with the use of "entanglement swapping" for teleportation [1] or
targeting observables of  a specific structure in order to avoid
decoherence [7]. Quantum entanglement relativity [2-6] and
relativity of the more general quantum correlations [10] open new
possibilities in manipulating the quantum information hardware. As
a matter of fact, we just start to learn about the physical
subtlety and importance of the concept of "quantum subsystem".

In this paper, we do not tackle the related deep questions.
Rather, as  a contribution to this new discourse in quantum
theory, we stick to a concrete model that can be solved
analytically and we provide some interesting observations.

We consider a pair of un-coupled modes in "phase space"
representation (as a pair of non-interacting linear harmonic
oscillators) independently subjected to the quantum amplitude
damping channels [13-16]. We analytically (exactly) solve the
Heisenberg equations of motion in the Kraus representation [13-19]
and analyze the results obtained for  the original as well as for
some alternate degrees of freedom. We find that the environment
non-equally "sees" the different structures. Particularly, {\it
only one structure is distinguished} by the locality of the
environmental influence on the structure's subsystems that
provides a classical-like description of the subsystems.

This paper is arranged as follows. In Section 2 we re-derive the
solutions to the Heisenberg equations for a pair of modes. Our
derivation is specific as it is an exact calculation in the {\it
infinite-sum} Kraus representation of the amplitude damping
dynamics of the two-mode system. In Section 3 we introduce and
analyze the alternate degrees of freedom (the alternate
structures) for the pair of modes and we obtain the Heisenberg
equations of motion for the new degrees of freedom. In Section 4
we emphasize the special characteristics of the original degrees
of freedom that do not apply to the alternate degrees of freedom.
Section 5 is conclusion.

\bigskip

{\bf 2. The model}

\bigskip

We consider the two uncoupled modes in the respective "phase
space" representations [16], i.e. as a pair of noninteracting
linear oscillators, $1$ and $2$, with the respective frequencies
and masses $\omega_1, \omega_2$. and $m_1, m_2$. The "phase space"
position variables, $x_1$ and $x_2$, and the conjugate momentums,
$p_1$ and $p_2$, respectively. The total Hilbert state space
factorizes $\mathcal{H} = \mathcal{H}_1 \otimes \mathcal{H}_2$ and
the total Hamiltonian $H = H_1 + H_2$, $H_i = p_i^2/2m_i +
m_i\omega_i^2 x_i^2/2$, $i=1,2$.

We assume the oscillators are independently subjected to the
quantum amplitude channels that can be described by mutually
independent master equations (in the interaction picture for zero
temperature) [14]:

\begin{eqnarray}
&\nonumber& \dot{\rho}_1 = \kappa_1 (2 a_1\rho_1 a^{\dag}_1 -
a^{\dag}_1 a_1 \rho_1 - \rho_1 a^{\dag}_1 a_1).
\\&&
\dot{\rho}_2 = \kappa_2 (2 a_2\rho_1 a^{\dag}_2 - a^{\dag}_2 a_2
\rho_2 - \rho_2 a^{\dag}_2 a_2),
\end{eqnarray}

\noindent with the respective "annihilation" ("creation")
operators $a_i$ ($a^{\dag}_i$, $i=1,2$) and, in general, with the
different damping parameters $\kappa_i, i=1,2$. The initial state
is tensor product, $\rho_{12}(0) = \rho_1(0) \otimes \rho_2(0)$.

The solutions to Eq. (1) are well-known  both in the Schr\"
odinger as well as in the Heisenberg picture, cf., e.g., Ref.
[16]. As we need the  observables (position and momentum
operators) and their bi-linear forms, below we re-derive the
respective expressions that facilitates further analysis.

The master equations Eq. (1) are known to be representable in the
Kraus form [13-16]:

\begin{equation}
\rho_i(t) = \sum_{n=0}^{\infty} K^i_n(t) \rho_i(0) K^{i\dag}_n(t),
\quad i=1,2
\end{equation}

\noindent with the completeness relation $\sum_{n=0}^{\infty}
K^{i\dag}_n(t) K^i_n(t) = I_i, i=1,2, \forall{t}$. For the
amplitude damping process, i.e. for the master equations Eq.(1),
the Kraus operators read [13-16]:

\begin{equation}
K^i_n (t) =  \sqrt{{(1-e^{-2\kappa_i t})^n \over n!}}
e^{-\kappa_itN_i} a_i^n; \quad N_i = a_i^{\dag} a_i, \quad i=1,2.
\end{equation}

In the Heisenberg picture, the states $\rho_i$ do not evolve in
time. Then dynamics is presented for every oscillator's observable
$A$ in the Kraus representation:

\begin{equation}
A_i(t) = \sum_{n=0}^{\infty} K^{i\dag}_n(t) A_i(0) K^{i}_n(t),
\quad i=1,2.
\end{equation}

Sometimes, the infinite sum in Eq.(4) is approximated by a few
first terms,  e.g. in Ref. [20]. However, below we give the {\it
exact} solutions to Eq.(4) without calling for or imposing any
approximation. As the two oscillators dynamics are mutually
independent, further on, we drop the index $i$ thus providing the
expressions relevant for both oscillators.

Substituting Eq. (3) into Eq. (4) one obtains:

\begin{equation}
A(t) = \sum_{n+0}^{\infty} {(1-e^{-2\kappa t})^n \over n!} a^{\dag
n} e^{-\kappa tN} A(0) e^{-\kappa tN} a^n.
\end{equation}

Bearing in mind:

\begin{equation}
x = \left({\hbar \over 2m\omega}\right)^{1/2} (a + a^{\dag}),
\quad p = \imath \left({m\hbar \omega \over 2}\right)^{1/2}
(a^{\dag} - a),
\end{equation}

\noindent we exchange $A(0)$ in Eq. (5) by $a$, $a^{\dag}$ and
$a^{\dag}a$.

To simplify the calculation, we make use of the following
generalization of the Baker-Hausdorff lemma [21]:

\begin{equation}
e^{-sA}Be^{-sA}=B-s\{A,B\}+\frac{s^2}{2!}\{A,\{A,B\}\}-\frac{s^3}{3!}\{A,\{A,\{A,B\}\}\}+....
\end{equation}

\noindent where the curly brackets denote the anticommutator,
$\{A, B\} = AB + BA$.

Then it is straightforward to obtain:

\begin{equation}
e^{-\kappa tN} a e^{-\kappa tN} = e^{-\kappa t} a e^{-2\kappa tN},
\quad e^{-\kappa tN} a^{\dag} e^{-\kappa tN} = e^{-\kappa t}
a^{\dag} e^{-2\kappa tN}.
\end{equation}

Returning Eq. (8) into the Heisenberg equations for $a$ and
$a^{\dag}$, one obtains:

\begin{equation}
A(t) = e^{\pm \kappa t} \sum_{n=0}^{\infty} {(1-e^{-2\kappa t})^n
\over n!} a^{\dag n} A(0) e^{-2\kappa tN} a^n
\end{equation}

\noindent with the positive exponent for $a$ and the negative
exponent for $a^{\dag}$.

For $A \equiv a^{\dag}$ one directly obtains:

\begin{equation}
a^{\dag}(t) = e^{-\kappa t} a^{\dag} \sum_{n=0}^{\infty}
{(1-e^{-2\kappa t})^n \over n!} a^{\dag n} e^{-2\kappa tN} a^n =
e^{-\kappa t} a^{\dag};
\end{equation}

\noindent the last equality in Eq. (10) follows from Eq. (8) and
from the completeness relation for the Kraus operators.

For $a(t)$ we obtain:

\begin{equation}
a(t) = e^{\kappa t}  \sum_{n=0}^{\infty} {(1-e^{-2\kappa t})^n
\over n!} [a^{\dag n}, a] e^{-2\kappa tN} a^n + e^{\kappa t} a
\sum_{n=0}^{\infty} {(1-e^{-2\kappa t})^n \over n!} a^{\dag n}
e^{-2\kappa tN} a^n
\end{equation}

\noindent where $[A, B] = AB - BA$ is the commutator.

With the aid of $[a^{\dag n}, a] = - n a^{\dag n-1}$, Eq. (11)
reads:

\begin{equation}
a(t) = - e^{\kappa t}  \sum_{n=0}^{\infty} n {(1-e^{-2\kappa t})^n
\over n!} a^{\dag n-1} e^{-2\kappa tN} a^n + e^{\kappa t} a.
\end{equation}

As it is easy to prove for the sum in Eq. (12):
$\sum_{n=0}^{\infty} n {(1-e^{-2\kappa t})^n \over n!} a^{\dag
n-1} e^{-2\kappa tN} a^n = (1-e^{-2\kappa t})a$, one finally
obtains:

\begin{equation}
a(t) = e^{-\kappa t}a.
\end{equation}

In completely the same way the terms $a^2(t)$, $a^{\dag  2}(t)$ as
well as $(a^{\dag}a) (t)$ can be calculated to obtain:

\begin{eqnarray}
&\nonumber& a^2(t) = \sum_{n=0}^{\infty} K^{\dag}_n(t) a^2 K_n(t)
= e^{-2\kappa t} a^2
\\&&\nonumber
a^{\dag 2}(t) = \sum_{n=0}^{\infty} K^{\dag}_n(t) a^{\dag 2}
K_n(t) = e^{-2\kappa t} a^{\dag 2}
\\&&
(a^{\dag}a)(t) = \sum_{n=0}^{\infty} K^{\dag}_n(t) a^{\dag}a
K_n(t) = e^{-2\kappa t} a^{\dag} a.
\end{eqnarray}

Now we can write the desired solutions to the Heisenberg equations
for the oscillator phase-space observables as follows:

\begin{eqnarray}
&\nonumber& x(t) = e^{-\kappa t} x, \quad p(t) = e^{-\kappa t} p,
\\&&\nonumber
x^2(t) = e^{-2\kappa t} x^2 + {\hbar \over 2 m\omega} (1 -
e^{-2\kappa t})
\\&&
p^2(t) = e^{-2\kappa t} p^2 + {m\hbar\omega \over 2 } (1 -
e^{-2\kappa t})
\end{eqnarray}

From Eq.(15) we directly obtain the asymptotic solutions:

\begin{equation}
\lim_{t\to\infty} x(t) = 0 = \lim_{t\to\infty} p(t),
\lim_{t\to\infty} x^2(t) =  {\hbar \over 2 m\omega},
\lim_{t\to\infty} p^2(t) =  {m\hbar\omega \over 2 },
\end{equation}

\noindent while Eq. (16) gives rise directly to:

\begin{equation}
\lim_{t\to\infty} \Delta x(t) \Delta p(t) = {\hbar \over 2};
\end{equation}

\noindent in Eq.(17) appear the standard deviations of the
respective observables.

Nonzero value of the covariance function, $C = \langle A_1 A_2
\rangle - \langle A_1 \rangle \langle A_2 \rangle$, reveals
correlations in the total $1+2$ system's state, $\rho_{12}$; the
observables $A_i, i=1,2$, refer to the two oscillators, while the
symbol $\langle \ast \rangle$ denotes the state averaging. In the
Heisenberg picture,  $\langle A_1(t) A_2(t) \rangle = tr_{12}
A_1(t)A_2(t)\rho_{12}(0)$. In general, $C=0$ does not guarantee
the absence of correlations. However, as we are interested in the
Gaussian states and in the (gaussianity-preserving) the amplitude
damping channel [13-19], the zero value of the covariance function
implies the absence of any correlations between the two modes.

Picking $A_1$ from the set $\{x_1, p_1\}$ and $A_2$ from the set
$\{x_2, p_2\}$, one can form the covariance functions $C_{ij}$ for
the different combinations. For the independent channels
considered here, one Kraus operator for the total system,
$K^1_m(t) \otimes K^2_n(t)$, with the completeness relation
$\sum_{m,n=0}^{\infty} K^{1\dag}_m(t) \otimes K^{2\dag}_n(t)
K^1_m(t) \otimes K^2_n(t) = I_{12}, \forall{t}$.

Now, with the definition

\begin{equation}
A_1(t) A_2(t) = \sum_{m,n=0}^{\infty} K^{1\dag}_m(t) A_1(0)
K^1_m(t) \otimes K^{2\dag}_n(t) A_2(0) K^2_n(t)
\end{equation}

\noindent and with the aid of  Eq. (15), one easily obtains:

\begin{equation}
\lim_{t\to\infty} C_{ij}(t) = \lim_{t\to\infty}
e^{-(\kappa_1+\kappa_2)t} C_{ij}(0) = 0, \quad \forall{i,j}.
\end{equation}

The expressions Eq.(16) and (19) directly provide the following
conclusion:  asymptotic state for the two-mode system is
tensor-product (cf. Eq. (19)) of the minimal-uncertainty (cf. Eq.
(17)) Gaussian states.

\bigskip

{\bf 3. The alternate degrees of freedom}

\bigskip

We introduce the alternate degrees of freedom, $X_A$ and $\xi_B$,
with the respective conjugate momentums, $P_A$ and $\pi_B$; $[X_A,
P_A] = \imath \hbar$, $[\xi_B, \pi_B] = \imath \hbar$. The formal
subsystems $A$ and $B$ define an alternate structure, $A+B$, for
the composite system $C\equiv1+2$. The total Hilbert state space,
$\mathcal{H} = \mathcal{H}_A \otimes \mathcal{H}_B$, while the
total Hamiltonian, $H = H_A + H_B + H_{AB}$.

Let us for simplicity consider the linear canonical
transformations (LCTs):

\begin{eqnarray}
&\nonumber& X_A = \sum_i \alpha_i x_i, P_A = \sum_j \gamma_j p_j
\\&&
\xi_B = \sum_m \beta_m x_m, \pi_B = \sum_n \delta_n p_n
\end{eqnarray}

\noindent that give rise to the constraints:

\begin{equation}
\sum_i \alpha_i \gamma_i = 1 = \sum_i \beta_i \delta_i, \quad
\sum_i \alpha_i \delta_i = 0 = \sum_i \beta_i \gamma_i.
\end{equation}

In general, the LCTs change the tensor-product factorization of
the total-system's Hilbert state space, the form of the
Hamiltonian (typically, the subsystems $A$ and $B$ interact to
each other) as well as the form of the total system's state [2, 7,
9]. As to the later, the absence of correlations (quantum or
classical) for e.g. the $1+2$ structure can give rise to some kind
of correlations for the $A+B$ structure--the recently observed
correlations relativity [9].

Despite the fact that the Kraus operator $K^1_m \otimes K^2_n$ is
composed of the local Kraus operators for the $1+2$ structure, due
to Eq. (20), this becomes {\it non-local} operation for the $A+B$
structure, $K^1_m \otimes K^2_n \neq K^A_p \otimes K^B_q$. In
other words: while there are two independent environments for the
$1$ and $2$ oscillators, the systems $A$ and $B$ {\it  share the
same environment}.

Of course, the composite system's state and the Hamiltonian are
unique in every instant in time for every possible structure.
Consequently, dynamics of the total system is described by unique
master equation for the total state $\rho_{12}$. However, for the
above distinguished reasons, the separation of the dynamics is not
in general possible for the alternate subsystems $A$ and $B$.
Deriving master equations for the subsystems $A$ and $B$ from the
master equations Eq.(1) is as yet largely intact. Rigorously, the
Kraus operators for the $A+B$ structure can follow only from the
master equation for the $A+B$ structure. Fortunately enough, these
Kraus operators are not necessary for our consideration. As
emphasized above: dynamics of the total system is uniquely defined
by the infinite sum of the form of Eq. (18), not by the particular
Kraus operators. Therefore, it suffices  to know the dynamics in
the terms of one structure--here of the $1+2$ structure, Eq. (18).
So, with the use of Eq. (15), we can still draw some conclusions
regarding the total system's state relative to the $A+B$
structure.

With the use of Eqs. (18) and  (20):

\begin{equation}
X_A(t) = \alpha_i x_i(t), \xi_B(t) = \beta_i x_i(t), P_A(t) =
\gamma_i p_i(t), \pi_B(t) = \delta_i p_i(t);
\end{equation}

\noindent in Eq.(22), we assume summation over the repeated
indices. On the other hand:

\begin{equation}
X_A^2(t) = \alpha_i \alpha_j (x_ix_j)(t), \quad P_A^2(t) =
\gamma_i \gamma_j (p_ip_j)(t),
\end{equation}

\noindent where Eq. (18) gives $(x_ix_j)(t) \equiv \sum_{m,n}
K^{1\dag}_m \otimes K^{2\dag}_n x_i x_j K^1_m \otimes K^2_n$; and
analogously for the subsystem $B$.

Eq. (22) and Eq. (23) give rise to the standard deviation:

\begin{eqnarray}
&\nonumber& (\Delta X_A(t))^2 = tr_{12} \alpha_i \alpha_j
(x_ix_j)(t) \rho_{12}(0) - (tr_{12} \alpha_i x_i(t)\rho_{12}(0))^2
=
\\&&\nonumber
\alpha_i \alpha_j \langle (x_ix_j)(t)\rangle - \alpha_i \alpha_j
\langle x_i(t)\rangle \langle x_j(t)\rangle =  \sum_i \alpha_i^2
(\Delta x_i(t))^2 + \\&& \sum_{i\neq j} \left(\alpha_i \alpha_j
\langle (x_ix_j)(t) \rangle - \alpha_i \alpha_j \langle
x_i(t)\rangle \langle x_j(t)\rangle\right) = \sum_i \alpha_i^2
(\Delta x_i(t))^2.
\end{eqnarray}

The last equality follows from the observation that, for $i\neq
j$, the locality of the Kraus operators for the $1+2$ structure,
Eq. (18), gives rise to the equality $\langle (x_ix_j)(t)\rangle =
\langle x_i(t)\rangle \langle x_j(t)\rangle$; remind: the initial
state $\rho_{12}(0) = \rho_1(0)\otimes \rho_2(0)$. In complete
analogy, one obtains the standard deviations for $P_A$, $\xi_B$
and $\pi_B$. Then easily follow the products of the asymptotic
standard deviations:

\begin{eqnarray}
&\nonumber& \Delta X_A(\infty) \Delta P_A(\infty) = \sqrt{\Large(
 {\alpha_1^2\hbar \over 2m_1 \omega_1} +  {\alpha_2^2\hbar \over
2m_2\omega_2}\Large) \Large(  {\gamma_1^2 m_1\hbar \omega_1 \over
2} +  {\gamma_2^2 m_2 \hbar\omega_2 \over 2}\Large)} \ge {\hbar
\over 2}
\\&&
\Delta \xi_B(\infty) \Delta \pi_B(\infty) = \sqrt{\Large(
{\beta_1^2\hbar \over 2m_1 \omega_1} +  {\beta_2^2\hbar \over
2m_2\omega_2}\Large) \Large(  {\delta_1^2 m_1\hbar \omega_1 \over
2} +  {\delta_2^2 m_2 \hbar\omega_2 \over 2}\Large)} \ge {\hbar
\over 2}
\end{eqnarray}

On the other hand, given the more general considerations [8], one
can expect that an alternative structure $A+B$ is described by
correlation of their subsystems, $A$ and $B$. This expectation can
here be tested by considering the covariance functions, e.g.:

\begin{equation}
C = \langle X_A(t) \xi_B(t) \rangle - \langle X_A(t) \rangle
\langle \xi_B(t)\rangle.
\end{equation}

\noindent Substituting Eqs.(15) and (22) into Eq.(26), one finds
in the asymptotic limit:

\begin{equation}
C = \alpha_i \beta_i (\Delta x_i(\infty))^2 = \alpha_1 \beta_1
{\hbar \over 2m_1\omega_1} + \alpha_2 \beta_2 {\hbar \over
2m_2\omega_2}.
\end{equation}

\bigskip

{\bf 4. Analysis and discussion of the results}

\bigskip

The model of Section 2 bears some important features. First, we do
not assume interaction between the two modes (oscillators). For
the coupled modes,  even Markovianity of the evolution requires
justification [19, 22]. Second, we assume the two independent
(mutually noninteracting) environments for the two  modes. While
this simplification is mathematically welcome [18, 19, 22], it is
physically remarkable: the total environment {\it locally} (i.e.
mutually independently) influences the two modes (oscillators).
This possibility to separate the total environment into two parts
locally "monitoring" the two modes (thus giving rise to Eq.(1))
does not apply to the alternate subsystems of the open system $C$.

The asymptotic quantum state for the $1+2$ structure is
"classical": it is tensor-product of a pair of the "coherent" (the
minimal-uncertainty Gaussian) states. The amplitude damping is a
CP map [17-19] and therefore cannot induce any (quantum or
classical) correlations for the input tensor-product states [23].
On the other hand, the "coherent" states are arguably the most
classical of all quantum states [16, 18, 19] (and the references
therein). These states are also known to represent the most robust
(the "preferred") states of a system described by the master
equation Eq.(1) [24]. Thereby, [in the asymptotic limit], one can
imagine the pair $1+2$ as a pair of "individual", mutually
distinguishable and non-correlated systems bearing the (robust)
quantum states of their own. In a sense, this is a definition of
"classical systems". So, one can say the environment composed of
the two noninteracting parts that provide mutually independent
(local) amplitude damping processes makes the $1+2$ structure
special.

It is worth repeating: the Kraus operators Eq.(3) are non-local
for any (non-trivial) alternate structure $A+B$. Interestingly
enough, a special choice of the parameters can provide the
classicality also for an alternate structure. For the resonant
oscillators ($\omega_1 = \omega_2 = \omega$) of equal masses ($m_1
= m_2 = m$), one obtains equalities in Eq. (25) and $C = 0$ in Eq.
(28) for the $A$ appearing as the center of mass of the two
oscillators ($\alpha_1 = 1/2 = \alpha_2$) and for the $B$
appearing as the "internal" degree of freedom ($\beta_1 = 1 = -
\beta_2$). This model, as  the simplest one possible for a pair of
oscillators, is often analyzed, cf. e.g. [8, 25, 26]. However,
already for the oscillators of non-equal masses [22], or of the
equal masses but non-resonant [27], this cannot be obtained.
Thereby and therefore, due to the local character of the {\it
environmental influence}, the classical-like structure $1+2$ is
special, i.e. distinguished in the set of the possible structures
of the composite system $C$.

In the more general context, our considerations exhibit usefulness
of the simple (or simplified) models: Analytic solutions for the
simple models  can still may serve for obtaining some information
about the more elaborate models, if the two are mutually linked
via the proper LCTs. The proper LCTs performed on the model (on
the structure) $1+2$  can introduce another pair of harmonic
oscillators, $A$ and $B$. The new oscillators can differ from the
original ones in a number of instances. E.g., not only their
respective masses and frequencies may differ, but (as distinct
from the original $1+2$) the new oscillators  share the same
environment and can still be in mutual interaction [2, 10]. If the
analytic solutions to the Heisenberg equations for the original
pair are known, one can easily obtain the analytic solutions to
the Heisenberg equations for the "new" oscillators. Therefore the
structure- (i.e. the LCTs-) based considerations provide a new
method for investigating the open bipartite systems and their
dynamics--just start from a simple model. To this end, the details
will be presented elsewhere.

Markovian dynamics is expected to provide the "classical"
description of an open system in the asymptotic limit [28]. To
this end, our considerations provide a new lesson: for the model
considered, the "classicality" is a {\it matter of a special
structure} of the composite system $C$. The  special structure is
chosen by the composite system's environment, and particularly by
the condition of local influence of the total environment on the
constituent subsystems.

This classical-like picture changes for the isolated composite
systems [9]. On the other hand, the alternate degrees of freedom
 of the open {\it quantum} systems may bear nontrivial physical interest and
 use. To this end, we refer the reader to the relevant literature
[2-12, 25, 27] (and the references therein).

\bigskip

{\bf 5. Conclusion}

\bigskip

It is a phenomenological fact: not all the observables (e.g. the
degrees of freedom) of a composite system are equally accessible
in a laboratory. With the use of a simple model of an open
two-mode system we show why it may be so. Our considerations
exhibit that, in the asymptotic limit, there is only one set of
the degrees of freedom that exhibits a classical-like description
as a consequence of a local environmental influence.

\bigskip

{\bf References}

\bigskip

[1] Bennett C H, Brassard G, Cr\' epeau C et al 1993 {\it Phys.
Rev. Lett.} {\bf 70} 1895

[2] Dugi\' c M and Jekni\' c J 2006  {\it Int. J. Theor. Phys.}
{\bf 45} 2249

 [3] Zanardi P 2001 {\it Phys. Rev. Lett.} {\bf 87}
077901

 [4] Ciancio E, Giorda P and Zanardi P 2006 {\it  Phys. Lett.} A {\bf
 354} 274

 [5] Barnum H, Knill E, Ortiz G, Somma R and Viola L 2003 {\it Phys. Rev. Lett.} {\bf 92} 107902

 [6] de la Torre A C, Goyeneche D and Leitao L 2010 {\it  Europ. J. Phys.} {\bf 31} 325

 [7] Jekni\' c-Dugi\' c J and Dugi\' c M 2008 {\it  Chin. Phys. Lett.} {\bf 25} 371

 [8] Chou C H, Yu T  and Hu B L 2008 {\it  Phys. Rev.} E {\bf 77} 011112

 [9] Dugi\' c M, Arsenijevi\' c M and Jekni\' c-Dugi\' c J 2012  {\it Sci. China-Phys. Mech.
 Astron.} (accepted)

 [10] Dugi\' c M and Jekni\' c-Dugi\' c J  2012 {\it Pramana} DOI:
10.1007/s12043-012-0296-3

 [11] Stokes A, Kurcz A, Spiller T P and Beige A 2012 {\it  Phys. Rev. A} {\bf 85} 053805

[12] Jekni\' c-Dugi\' c J,  Dugi\' c M, Francom A and Arsenijevi\'
c M 2012 E-print arXiv:1204.3172

[13] Fan H Y and Hu L Y 2009 {\it Opt. Commun.} {\bf 282} 932

[14] Jiang N Q, Fan H Y, Xi L S, Tang L Y and Yuan X Z 2011 {\it
Chin. Phys.} B {\bf 20} 120302

[15] Zhou N R, Hu L Y and Fan H Y 2011 {\it  Chin. Phys.} B {\bf
20} 120301

[16] Ferraro A, Olivares S and  Paris M G A 2005 {\it Gaussian
States in Quantum Information} (Napoli: Bibliopolis)

[17] Kraus K 1983 {\it States Effects and Operations Fundamental
Notions of Quantum Theory, Lecture Notes in Physics} Vol. 190
(Berlin: Springer-Verlag)

[18] Breuer H P and   Petruccione F 2002 {\it The Theory of Open
Quantum Systems} (Oxford: Clarendon Press)

[19] Rivas A and Huelga S F 2011 {\it Open Quantum Systems: An
Introduction} (Berlin: Springer)

[20] Liu Y X, Ozdemir S K, Miranowicz A and Imoto N 2004 {\it
Phys. Rev.} A {\bf 70} 042308

[21] Menda\v s I P and Popovi\' c D B 2010 {\it Phys. Scripta}
{\bf 82} 045007

[22] Rivas A, Plato A D K, Huelga S F and Plenio M B 2010 {\it New
J. Phys.} {\bf 12} 113032

[23] Arsenijevi\' c M, Jekni\' c-Dugi\' c J and Dugi\' c M 2012
E-print arXiv:1204.2789

[24] Boixo S, Viola L and Ortiz G 2007 {\it  Europhys. Lett.} {\bf
79} 40003

[25] Freitas J N and Paz J P 2012 {\it Phys. Rev.} A {\bf 85}
032118

[26] Liu K L and Goan H S 2007 {\it Phys. Rev.} A {\bf 76} 022312

[27] Paz J P and Roncaglia A J 2008 {\it  Phys. Rev. Lett.} {\bf
100} 220401

[28] Ferraro A, Aolita L, Cavalcanti D, Cucchietti F M and  Acin A
2010 {\it Phys. Rev.} A {\bf 81} 052318

\end{document}